\documentclass{appolb}
\usepackage{epsfig}
\begin{document}
%\date{\today}
\newcommand{\imag}{\Im {\rm m}}
\newcommand{\real}{\Re {\rm e}}

\pagestyle{plain}
%% uncomment the following line to get equations numbered by (sec.num)
%\eqsec
\newcount\eLiNe\eLiNe=\inputlineno\advance\eLiNe by -1
\title{\hfill IFT-3/011\\ \phantom{}\\
CP violation in the neutralino system
\thanks{Dedicated to Jan Kwieci\'nski  on his 65th birthday.}%
}
\author{Jan KALINOWSKI
\address{Instytut Fizyki Teoretycznej, Uniwersytet Warszawski\\
Ho\.za~69,  00-681~Warszawa,  Poland}}
\maketitle

\begin{abstract}
Supersymmetric extensions of the Standard Model provide new sources of
CP violation.   Here the CP properties of neutralinos are described 
and possible experimental signatures of CP--violation in the
neutralino production processes at $e^+e^-$ linear colliders are
discussed.

\end{abstract}
%\pacs{PACS number(s): 11.30.Er, 12.60.Jv, 13.10.+q}
%\noindent IFT-3/011
%\newpage

%%%%%%%%%%%%%%%%%%%%%%%%%%%%%%%%%%%%
\section{Introduction}
%%%%%%%%%%%%%%%%%%%%%%%%%%%%%%%%%%%%
The electroweak sector of the Standard Model (SM) contains only one
CP-violating phase which arises in the Cabibbo-Kobayashi-Maskawa (CKM)
quark mixing matrix. Adding right-handed neutrinos to account for
non--zero neutrino masses and their mixing opens up a possibility of
new CP-violating phases in the Maki-Nakagawa-Sakata (MNS) lepton 
mixing matrix. In both cases unitarity imposes constraints on mixing
matrices, which can be represented graphically as unitarity
triangles. After many years of experimentation with the $K$ and $B$
mesons, the CKM unitarity triangle $V_{ud}V^*_{ub}+ V_{cd}V^*_{cb} +
V_{td}V^*_{tb}=0$ is essentially reconstructed and the required amount
of CP violation can be accommodated within the SM. (Reconstructing MNS
triangles will be far more difficult.) The CP violation in the SM also
induces electric dipole moments (EDM) of elementary
particles. However, in the lepton sector\footnote{An additional CP
violation due to strong interactions of the form
$\frac{\alpha_s}{8\pi}\theta G \tilde G$ could generate a very large
neutron EDM, unless $\theta$ is tuned to be very small.} they are
induced at a multiloop level and, as a result, EDM's generated by the
CKM phase are extremely small \cite{SMEDM} and beyond current and
foreseeable future experiments. Therefore the observation of a lepton
EDM would be a clear signal of new physics beyond the Standard Model.
 
While the CP properties of  the $K$ and $B$ systems appear to be
consistent with the SM, another (indirect)  
piece of evidence of CP violation, the baryon asymmetry in
the universe,  requires a new
source of CP violation beyond what is in the SM \cite{genesis}. 
Thus new CP-violating phases must exist in nature.

Supersymmetric extensions of the SM based on soft supersymmetry
breaking mechanism introduce a plethora of CP phases. 
This poses a
SUSY CP problem (assuming even that the strong CP is solved), since if
the phases are large ${\cal O}(1)$, 
SUSY contributions to the lepton EDM
can be too large to satisfy current experimental constraints
\cite{susycp}.   
Many models have been proposed \cite{overcome} 
to overcome this problem: fine tune
phases to be small, push sparticle spectra above a few  TeV to suppress
effects of large phases on the EDM, constrain phases present in the
first two generations to be small, arrange for internal cancellations
etc. 

In the absence of any reliable theory that forces in a natural way the
phases to be vanishing or small, it is mandatory to consider scenarios with  
some of the phases large and 
arranged consistent with experimental EDM data. 
In such scenarios many
phenomena will be affected: sparticle masses, their decay rates and 
production cross sections, SUSY contributions to SM processes etc. 
It is one of the main physics goals of future collider
experiments to find SUSY and verify its CP properties \cite{futurecoll}.  
Detailed analyses of the neutralino sector can prove particularly
useful in this respect. In this note we will discuss the CP properties of
neutralinos strengthening an  argument of ref.\cite{ckmz}  that   
measurements of the neutralino production
cross sections  can provide a qualitative, unambiguous  evidence for
non--trivial CP phases.

\section{Neutralino sector of the MSSM}
In the minimal  supersymmetric extension of the Standard Model (MSSM),
the mass matrix of the  spin-1/2 partners of the neutral gauge 
bosons, $\tilde{B}$ and $\tilde{W}^3$, 
and of the neutral Higgs bosons, $\tilde H_1^0$ and 
$\tilde H_2^0$, takes the form   
\begin{eqnarray}
{\cal M}=\left(\begin{array}{cccc}
  M_1       &      0          &  -m_Z c_\beta s_W  & m_Z s_\beta s_W \\[2mm]
   0        &     M_2         &   m_Z c_\beta c_W  & -m_Z s_\beta c_W\\[2mm]
-m_Z c_\beta s_W & m_Z c_\beta c_W &       0       &     -\mu        \\[2mm]
 m_Z s_\beta s_W &-m_Z s_\beta c_W &     -\mu      &       0
                  \end{array}\right)\
\label{eq:massmatrix}
\end{eqnarray}
Here $M_1$ and $M_2$ are the fundamental supersymmetry breaking
parameters: the U(1) and SU(2) gaugino masses, and $\mu$ is the
higgsino mass parameter. As a result  of the electroweak symmetry
breaking by
the vacuum expectation values of the two neutral Higgs fields $v_1$
and $v_2$ ($s_\beta =\sin\beta$, $c_\beta=\cos\beta$ where
$\tan\beta=v_2/v_1$), non--diagonal terms with $s_W=\sin\theta_W$ and
$c_W=\cos\theta_W$ appear and  gauginos and
higgsinos mix to form the neutralino mass eigenstates $\chi_i^0$
($i$=1,2,3,4).

In general the mass parameters $M_1$, $M_2$ and $\mu$ in the mass matrix
(\ref{eq:massmatrix}) can be 
complex.  By reparameterization of the fields, $M_2$
can be taken real and positive; the
two remaining non--trivial phases, which are therefore 
reparameterization--invariant, may be attributed to $M_1$
and $\mu$:
\begin{eqnarray}
M_1=|M_1|\,\,{\rm e}^{i\Phi_1}\ \quad {\rm and} \qquad   
\mu=|\mu|\,\,{\rm e}^{i\Phi_\mu} \quad (0\leq \Phi_1,\Phi_\mu< 2\pi)
\end{eqnarray}

Since the existence of CP--violating
phases in supersymmetric theories in general induces electric dipole
moments (EDM),  current experimental bounds
can be 
exploited to derive indirect limits on the parameter space
\cite{susycp}. In fact  the
experimental limits on EDM's of the
electron, neutron and mercury atom have been used to (partly) justify  the
assumption of real SUSY parameters, and  most phenomenological studies
on supersymmetric particle   
searches have been performed within the CP-conserving MSSM.  
However, the EDM constraints can be avoided assuming
masses of the first and second generation sfermions large (above the
TeV scale), or arranging cancellations between the different SUSY
contributions to the EDMs. As a result, the complex phase
of the higgsino mass parameter $\mu$ is much less restricted than
previously assumed, while the complex phase of $M_1$ is practically
unconstrained. The possibility of non--zero CP--phases should therefore
be included in phenomenological analyses. 

The neutralino mass eigenvalues $m_i\equiv m_{\tilde{\chi}^0_i}$
$(i=1,2,3,4)$ can  be chosen
{\it positive} by a suitable definition of the unitary mixing matrix $N$. 
In  general this matrix involves 
6 angles and 10 phases, and can be written as \cite{ckmz,six}
\begin{equation}
N= {\sf diag}\left\{{\rm e}^{i\alpha_1},\, 
                   {\rm e}^{i\alpha_2},\,
                   {\rm e}^{i\alpha_3},\,
                   {\rm e}^{i\alpha_4}\,\right\} {\sf R}_{34}\, {\sf
                   R}_{24}\,{\sf R}_{14}\,{\sf R}_{23}\,{\sf
                   R}_{13}\, {\sf R}_{12}  
\label{eq:Mdef} 
%\qquad (0\leq \alpha_i < \pi {\rm  ~~mod~~} \pi)
\end{equation}
where ${\sf R}_{jk}$ are rotations in 
the [$jk$] plane characterized by a mixing angle $\theta_{jk}$ and a
(Dirac) phase $\beta_{jk}$. For example, 
\begin{eqnarray}
{\sf R}_{12}=\left(\begin{array}{cccc}
             c_{12}  &  s^*_{12}  &  0  &  0 \\
            -s_{12}  &  c_{12}    &  0  &  0 \\
                0    &      0     &  1  &  0  \\
                0    &      0     &  0  &  1 
                  \end{array}\right)
\end{eqnarray}
with $ c_{jk}\equiv \cos\theta_{jk}$, 
$s_{jk}\equiv \sin\theta_{jk}\, {\rm e}^{i\beta_{jk}}$.
One of (Majorana) phases $\alpha_i$ is nonphysical  and, for example, 
$\alpha_1$  may be chosen to vanish. %
None of the remaining 9  phases can be 
removed by rotating the fields 
since neutralinos are Majorana fermions.

Neutralino sector is CP conserving if $\mu$ and $M_1$ are real, which
is equivalent to  $\beta_{ij}=0$ (mod $\pi$) and
$\alpha_i=0$ (mod $\pi/2$). Majorana phases  $\alpha_i=\pm
\pi/2$  indicate only 
different CP parities of the neutralino states \cite{R2}.

The  matrix elements of $N$  define the
couplings of the mass eigenstates $\tilde{\chi}^0_i$ to other particles.
Like in the quark sector, it is useful \cite{ckmz,AASB} to represent the 
unitarity constraints on the elements $N_{ij}$
\begin{eqnarray}
 \label{eq:M}
M_{ij}&=& N_{i1}N^*_{j1}+N_{i2} N^*_{j2} + N_{i3} N^*_{j3}
               +N_{i4}N^*_{j4}=\delta_{ij} \\
D_{ij}&=& N_{1i}N^*_{1j}+N_{2i} N^*_{2j} + N_{3i} N^*_{3j}
               +N_{4i}N^*_{4j}=\delta_{ij} 
 \label{eq:D}
\end{eqnarray}
in terms of unitarity quadrangles. For $i\neq j$, the above equations
define the $M$- and $D$-type quadrangles in the complex plane. The
$M$-type quadrangles are formed by the sides $N_{ik}N^*_{jk}$
connecting two rows $i$ and $j$, and the $D$-type by
$N_{ki}N^*_{kj}$
connecting two columns $i$ and $j$ of the mixing matrix.
By a proper ordering of sides the quadrangles are assumed to be convex
with areas given by
\begin{eqnarray}
{\rm area}[M_{ij}]&=&{\textstyle \frac{1}{4}}
         (|J_{ij}^{12}|+|J_{ij}^{23}|+|J_{ij}^{34}|+|J_{ij}^{41}|)
\label{eq:aM}\\
{\rm area}[D_{ij}]&=&{\textstyle \frac{1}{4}}
         (|J_{12}^{ij}|+|J_{23}^{ij}|+|J_{34}^{ij}|+|J_{41}^{ij}|)
\label{eq:aD}
\end{eqnarray}
where $J_{ij}^{kl}$ are the Jarlskog--type CP--odd ``plaquettes''
\cite{CJ} 
\begin{equation}
J_{ij}^{kl}=\imag N_{ik}N_{jl}N_{jk}^*N_{il}^*
\label{eq:plaq}
\end{equation}
Note that plaquettes, and therefore the areas of unitarity
quadrangles, are not sensitive to the Majorana phases $\alpha_i$.

Unlike in the quark or lepton sector, the orientation of all
quadrangles is physically meaningful, and determined by the CP-phases
of the neutralino mass matrix.   
In terms of quadrangles, CP is conserved if and only if all 
quadrangles have null area (collapse to lines or points) {\it and} are
oriented along either real or imaginary axis.

\section{Experimental signatures of CP violation}

In principle, the imaginary parts of
the complex parameters involved could most directly and unambiguously
be determined by measuring suitable $CP$ violating observables.

Neutralinos can copiously be produced at prospective $e^+e^-$ linear
colliders \cite{futurecoll} via the $s$-channel $Z$ exchange and $t$- and
$u$-channel selectron exchange.   The
polarized differential cross section for the $\tilde{\chi}^0_i
\tilde{\chi}^0_j$ production is given by \cite{ckmz}
\begin{eqnarray}
&& \frac{{\rm d}\sigma^{\{ij\}}}{{\rm d}\cos\theta \,{\rm d}\phi}
  =\frac{\alpha^2}{16\, s}\, \lambda^{1/2} \bigg[
     (1-P_L\bar{P}_L)\,\Sigma_U+(P_L-\bar{P}_L)\,\Sigma_L
     \nonumber\\
&& { }\hskip 2.5cm 
  +P_T\bar{P}_T\cos(2\phi-\eta)\,\Sigma_T
  +P_T\bar{P}_T\sin(2\phi-\eta)\,\Sigma_N\bigg]\ \label{eq:diffx}
\end{eqnarray}
where $P$=$(P_T,0,P_L)$ [$\bar{P}$=$(\bar{P}_T \cos\eta,\bar{P}_T\sin\eta,
-\bar{P}_L)$] is the electron
[positron]  polarization vector; 
the electron--momentum direction
defines  the $z$--axis and
the electron transverse polarization--vector the $x$--axis; 
$\lambda=[1-(\mu_i+\mu_j)^2][1-(\mu_i-\mu_j)^2]$ 
with $\mu_i=m_i/\sqrt{s}$. 
The coefficients $\Sigma_U$, $\Sigma_L$, $\Sigma_T$
and $\Sigma_N$ depend 
only on the polar angle $\theta$ and their explicit form 
is given in \cite{ckmz}. 

An interesting feature of neutralino production is encoded in the term  
$\Sigma_N$ of eq.(\ref{eq:diffx}). Unlike $\Sigma_U$, $\Sigma_L$ and
$\Sigma_T$, the $\Sigma_N$  is a function of plaquettes 
\begin{eqnarray}
&& \Sigma_{N}=4\lambda \sin^2\theta \,\left[A_1 (J^{31}_{ij}-J^{41}_{ij})
          -A_2 (\tilde{J}^{32}_{ij}-\tilde{J}^{42}_{ij}) +A_3
\tilde{J}^{21}_{ij} \, \right] \label{sigmaN}
\end{eqnarray}
where tilde means that in calculating $J$  the $N_{i2}$ 
should be replaced by $N'_{i2}=s_W N_{i1}+c_W N_{i2}$. The
combinations of propagator 
factors $A_i$ are 
\begin{eqnarray}
&& A_1
   =\frac{1}{ 4c_W^4} D_Z(D_{tL}-D_{uL})\nonumber\\
&& A_2
   = \frac{s_W^2-1/2}{16s_W^4 c_W^4} D_Z
(D_{tR}-D_{uR})
                  \nonumber\\
&& A_3 
   =\frac{1}{8 s^2_W c^4_W} (D_{tL}D_{uR}-D_{tR}D_{uL})
\label{eq:normal1}
\end{eqnarray}
with $D_{tL,R}=s/(t-m^2_{\tilde{e}_{L,R}})$,
$D_{uL,R}=s/(u-m^2_{\tilde{e}_{L,R}})$ and 
the $Z$-boson propagator $D_Z=s/(s-m^2_Z)$ is taken real by
neglecting the $Z$ width in the limit of high energies.
Therefore $\Sigma_N$ is nonvanishing only in CP--noninvariant theories and 
already the detailed measurement of the angular
distribution of produced neutralino pairs in collisions of 
transverse polarized beams could indicate  the presence of CP phases.  

However,  since
$\Sigma_N$ depends on plaquettes, nonvanishing $\Sigma_N$ requires
specific form of CP-violation: the area of some unitarity 
quadrangles has to be non--zero, {\it i.e.} at least one of Dirac
phases $\beta_{kl}\neq 0$. Moreover, the effect might be quite small 
due to cancellations 
between $\tilde{H}^0_1$ and $\tilde{H}^0_2$ 
higgsino components 
%gaugino and higgsino components 
of $\tilde{\chi}^0_i$ and $\tilde{\chi}^0_j$ in eq.(\ref{sigmaN}), and
between t- and u-channel selectron exchanges in eq.(\ref{eq:normal1}).

If the initial beams are not polarized, the CP phases 
could be inferred from the ${\cal P}_N$ component of the 
polarization of the 
$\tilde{\chi}^0_i\tilde{\chi}^0_j$ pairs produced in $e^+e^-$ 
annihilation \cite{ckmz,Gudi}. 
The polarization vector $\vec{\cal P}=({\cal P}_L, {\cal P}_T, {\cal
P}_N) $  is defined in the rest frame of the particle
$\tilde{\chi}^0_i$, with components parallel to
the $\tilde{\chi}^0_i$ flight direction in the c.m. frame, in the
production plane, and normal to the production plane, respectively.
The normal component ${\cal P}_N$ can only be
generated by complex production amplitudes in the 
non--diagonal $\tilde{\chi}^0_i
\tilde{\chi}^0_j$ pair production process with $i\neq j$. 
For example, the contribution to ${\cal P}_N$ 
from the $\tilde{e}_R$ exchange
reads   
{\small
\begin{eqnarray}
&& {\cal P}_N =\frac{8\lambda^{1/2}\mu_j\,\sin\theta}{c^2_W \Sigma_{U}}
D_{uR}D_{tR} \imag\left[\, (N_{i1}N^*_{j1})^2\, \right]
\end{eqnarray}}
The normal polarization  can be non--zero even
if all the $\beta$-type CP phases vanish, {\it i.e.} it could 
signal the existence of
non--trivial $\alpha$-type CP phases. 

\begin{figure}[tb]
\begin{center}
  \epsfig{file=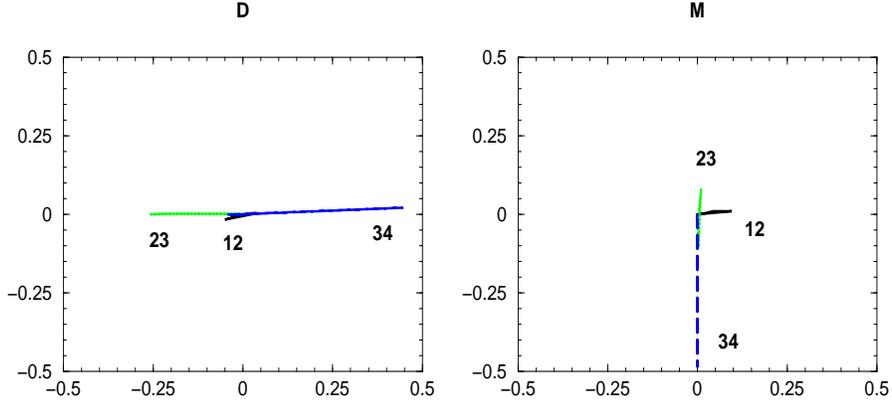,height=12cm,width=6cm,angle=-90} 
%  \vspace*{0.5cm}
\caption{\it The $D$--type (left panel) and $M$--type (right panel) 
  quadrangles in the complex plane,
  illustrated for $\tan\beta=10$, $|M_1|=100.5$ GeV, $\Phi_1=\pi/5$,
  $M_2=190.8$ GeV, $|\mu|=365.1$ GeV and $\Phi_{\mu}=0$; 
  $ij$ as indicated in the figure.}
\label{fig:quad}
\end{center}
\end{figure}

The non--zero values of CP-odd characteristics 
$\Sigma_N$ or ${\cal P}_N$ would unambiguously indicate CP-violation
in the neutralino sector. However, their experimental measurements 
will be  very difficult. On the other hand, one can also try to
identify the presence of 
CP-phases by studying their impact on the CP-even quantities, like
neutralino masses, branching ratios etc. Since these quantities are
non--zero in CP conserving case, the detection of CP-odd phases will
require a careful 
quantitative analysis of a number of physical observables. In
particular, for 
numerically small CP-odd phases, their deviations from CP-even values  
will also be small. As an example, in Fig.\ref{fig:quad} the unitarity
quadrangles for a particular point in the parameter space 
(consistent with
all experimental constraints) are
shown. The phase of $\mu$ is set to zero, and $\Phi_1=\pi/5$. 
In this case the quadrangles are almost degenerate to lines parallel
to either real or imaginary axis, and revealing the phase of $M_1$
will be quite difficult. 
%\vspace*{-0.5cm}

In this respect, as pointed out in ref.\cite{ckmz}, 
a clear indication of non--zero CP violating phases can be provided by
studying the energy behavior of the cross sections for  non--diagonal
neutralino pair production near thresholds. 

In CP-invariant theories, the CP parity of a pair of Majorana fermions
$\tilde{\chi}^0_i\tilde{\chi}^0_j$ is given by
\begin{eqnarray}
\eta=\eta^i\eta^j (-1)^L
\label{cpparity}
\end{eqnarray}
where $\eta^i$ is the CP parity of $\tilde{\chi}^0_i$ and $L$ is the
angular momentum \cite{11a}.  
Therefore neutralinos with the same CP parities (for
example for $i=j$) can be excited only in the P-wave via  
the s-channel $\gamma$ and $Z$ production processes. The
excitation in the S-wave, with the characteristic   
steep rise $\sim \lambda^{1/2}$ of the cross section near threshold,
can occur only for 
non--diagonal pairs with opposite CP--parities of the produced
neutralinos \cite{R2}.

The power of the selection rule (\ref{cpparity}) 
can clearly be seen by inspecting the expressions for 
the 
total cross section $\sigma^{\{ij\}}$ ($i\neq j$) near threshold 
\begin{figure}[tb]
\begin{center}
\epsfig{file=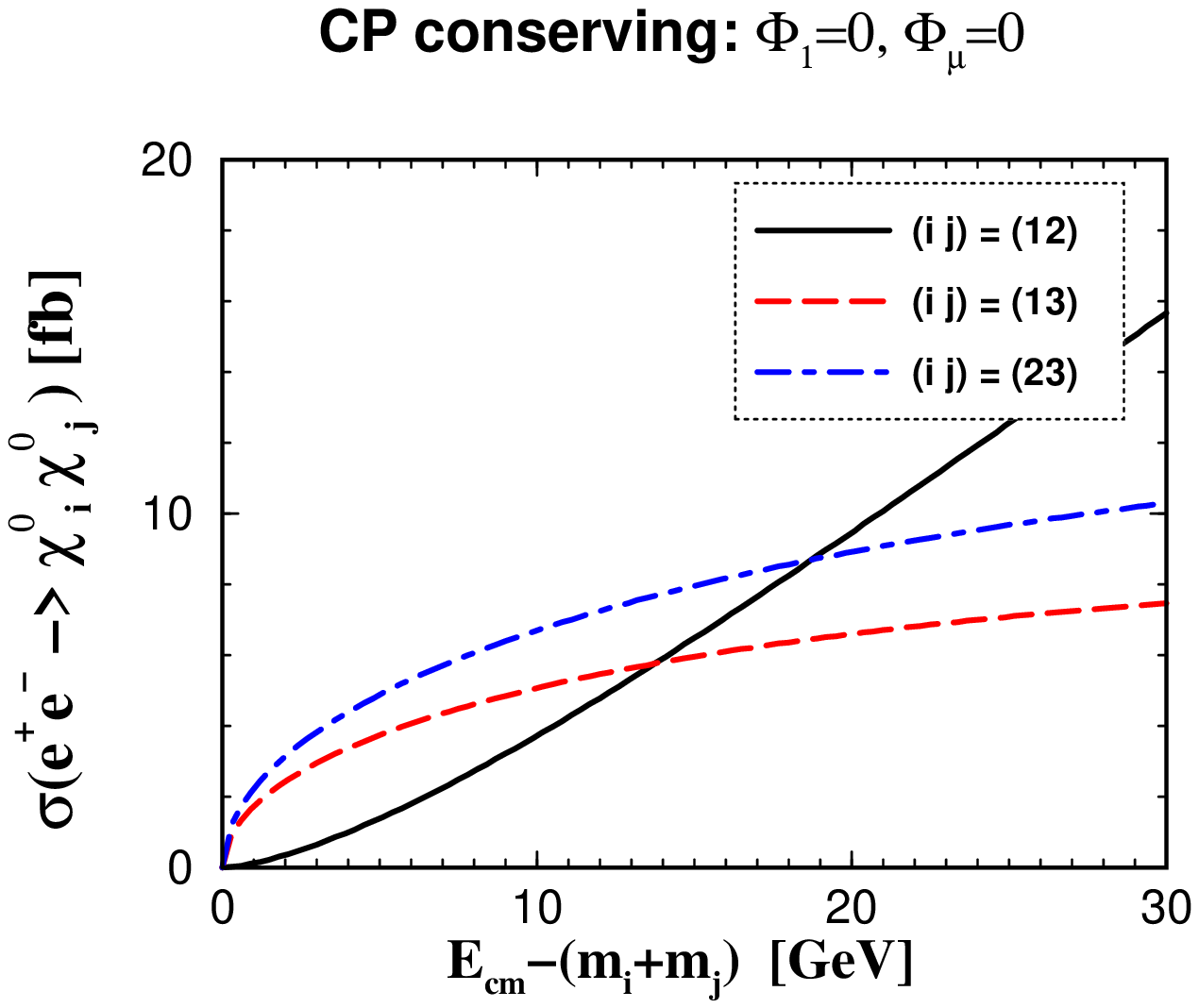,height=6cm,width=6cm} 
\epsfig{file=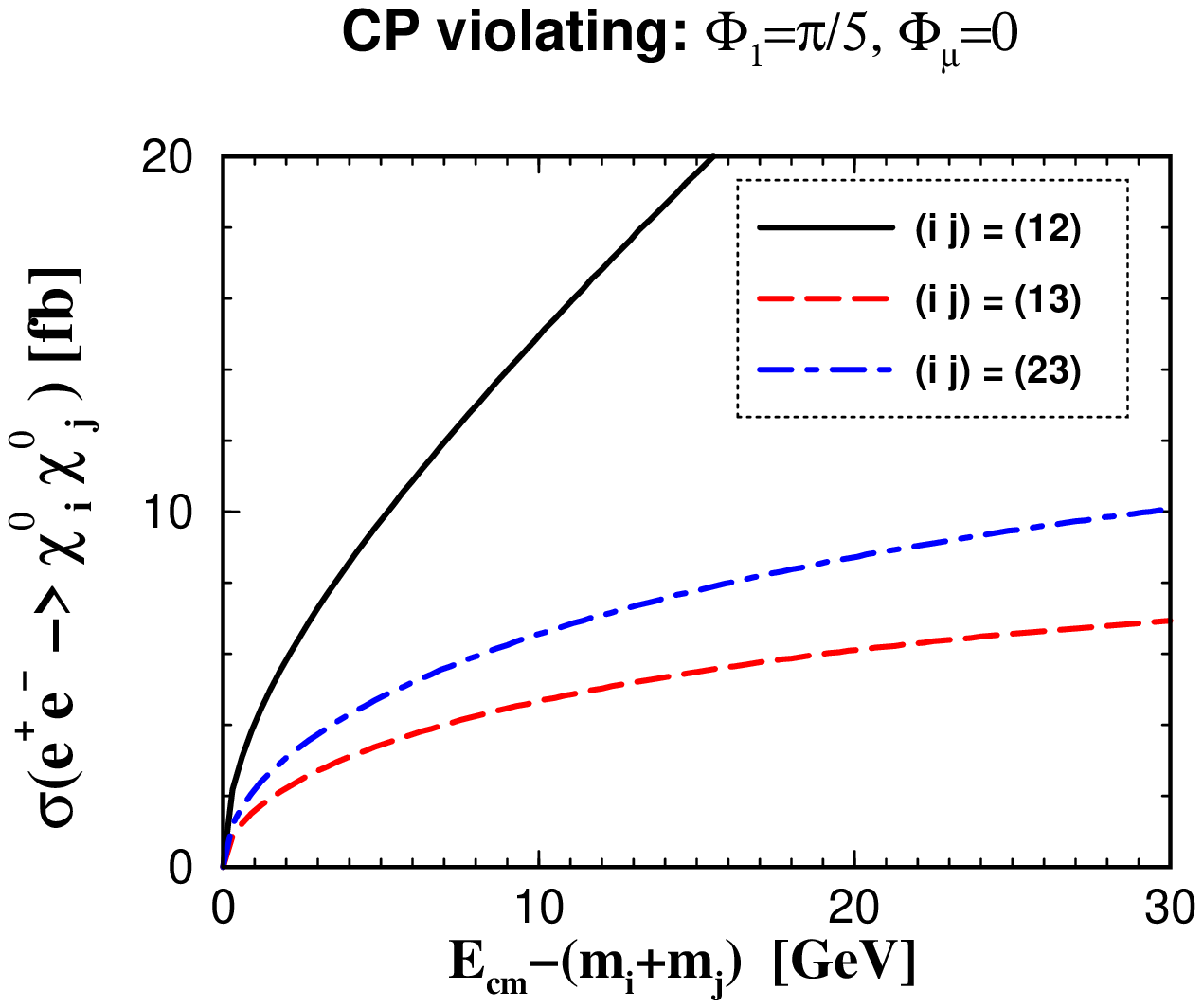,height=6cm,width=6cm} 
\caption{\it The threshold behavior of the neutralino  production
  cross--sections $\sigma^{\{ij\}}$ for the CP--conserving (left panel)
and the CP--violating (right panel) cases. Other parameters as in Fig.1.}
\label{fig:th}
\end{center}
\end{figure}
\begin{eqnarray}
\sigma^{\{ij\}} \approx \frac{\pi\alpha^2\,\lambda^{1/2}\,}{(m_i+m_j)^2}
          \,\bigg\{\, \frac{4m_im_j}{(m_i+m_j)^2}\,( |\imag\,
          G^{(0)}_R|^2\,
         +  |\imag\, G^{(0)}_L|^2) +
{\cal O}( \lambda)\, \bigg\}  
\label{eq:thres}
\end{eqnarray}
where  
\begin{eqnarray}
&& G^{(0)}_R=\frac{1}{2c_W^2}\, D_{0} (N_{i3}N^*_{j3}-N_{i4}N^*_{j4})
  - \frac{1}{c_W^2}F_{R} N_{i1}N^*_{j1}\nonumber \\
&& G^{(0)}_L=\frac{(s^2_W-1/2)}{2c_W^2 s_W^2}\, D_{0}
(N_{i3}N^*_{j3}-N_{i4}N^*_{j4})
+ \frac{1}{4s^2_Wc^2_W} F_{L}N'_{i2}N^{'*}_{j2}
%\nonumber \\ 
%&&{}\hspace{1cm}   + \frac{1}{4s^2_Wc^2_W} F_{L}(s_W N_{i1}+c_W N_{i2})(s_W
%N^*_{j1}+c_W N^*_{j2})  
\nonumber 
\end{eqnarray}
and the kinematic functions
\begin{eqnarray}
&& D_0=(m_i+m_j)^2/((m_i+m_j)^2-m_Z^2)\nonumber \\
&& F_{L,R}=(m_i+m_j)^2/(m^2_{\tilde{e}_{L,R}}+m_i m_j)\nonumber 
\end{eqnarray}
In the CP--invariant theory, the imaginary parts of  
$N_{ij}$ can only be generated by 
Majorana phases $\alpha_i=0$ and $\alpha_j=\pi/2$ or vice versa, {\it
i.e.} the S--wave excitation is possible when 
the CP--parities of the produced neutralinos are opposite, as dictated
by the eq.(\ref{cpparity}). 
This immediately implies that if 
the $\{ij\}$ and $\{ik\}$ pairs are excited in the S--wave, 
the pair $\{jk\}$ must be excited 
in the  P--wave characterized  by the
slow rise $\sim\lambda^{3/2}$ of the cross section, Fig.\ref{fig:th},
left panel. 

If, however, CP is violated 
the angular momentum of the produced neutralino
pair is no longer  restricted by the eq.(\ref{cpparity})  and 
all non--diagonal pairs can be excited in the  S--wave.
This is illustrated in Fig.\ref{fig:th}, where 
the threshold behavior of the neutralino pairs $\{12\}$, $\{13\}$ and
$\{23\}$ for the CP-conserving (left panel) case is contrasted to the
CP-violating case (right panel). 
Even for relatively small CP--phase $\Phi_1=\pi/5$,    
implying small impact on CP--even quantities,  
the change in the energy dependence near threshold can be quite
dramatic. 
Thus, observing the   $\{ij\}$,
$\{ik\}$ and $\{jk\}$ pairs to be excited {\it all} in S--wave states would 
therefore signal CP--violation.  
\section{Conclusions}
The supersymmetric extension of the Standard Model can come with new
sources of CP violation. In the absence of natural 
suppression of the SUSY  CP--phases, their non--zero values have to be
considered in phenomenological studies.   In this paper we have
discussed the CP properties of neutralinos, which are quite peculiar
due to their Majorana nature.

The CP violation in the neutralino sector can reveal itself in many
different ways. The most ambitious analysis would require the
experimental reconstruction of the unitarity quadrangles. Since all
phases of the mixing matrix $N$ (at least at the tree level) are
ultimately determined by the phases of the fundamental parameters
$\mu$ and $M_1$,  the reconstruction of the quadrangles would provide
many consistency checks of the underlying theory.

On the other hand, the first qualitative indication of the 
CP violation can be provided by the energy dependence of the
neutralino production cross sections. The steep rise of cross sections for
the production of at least three different non--diagonal neutralino
pairs can be interpreted as  a first direct 
signature  of the
presence of CP--violation in the neutralino sector.\\

\noindent {\bf Acknowledgments}\\
\noindent Work supported in part by the KBN Grant 2 P03B 040 24 (2003-2005).

\end{document}